\documentclass[acmsmall,screen]{acmart}

\AtBeginDocument{%
  \providecommand\BibTeX{{%
    \normalfont B\kern-0.5em{\scshape i\kern-0.25em b}\kern-0.8em\TeX}}}

\setcopyright{acmcopyright}
\copyrightyear{2024}
\acmYear{2024}
\acmDOI{10.1145/1122445.1122456}

\acmJournal{JACM}
\acmVolume{1}
\acmNumber{2}
\acmArticle{3}
\acmMonth{5}

\usepackage{color}

\usepackage{amssymb}
\usepackage{booktabs}
\usepackage[linesnumbered,ruled]{algorithm2e}
\usepackage{amsmath}
\usepackage{subfigure}
\usepackage{svg}

\usepackage{url}
\usepackage{algpseudocode}

\usepackage{threeparttable}     
\usepackage{multirow}
\usepackage{subfigure}
\usepackage{float}
\usepackage{makecell}
\usepackage{tabularx} 
\usepackage{graphicx}
\usepackage{xspace}
\usepackage{tcolorbox}
\usepackage{tikz}
\newcommand*{\circled}[1]{\lower.7ex\hbox{\tikz\draw (0pt, 0pt)%
    circle (.45em) node {\makebox[0.6em][c]{\small #1}};}}

\usepackage{booktabs}
\usepackage{amsmath}
\usepackage{amsmath}
\usepackage{array}
\usepackage{multirow}%
\usepackage{adjustbox} %
\usepackage{caption} %

\begin{document}

\title{Detecting the Root Cause Code Lines in Bug-Fixing Commits by Heterogeneous Graph Learning}


\author{Liguo~Ji}
\email{jgy1@dlmu.edu.cn}
\affiliation{%
  \institution{Dalian Maritime University}
   \country{China}
}

\author{Chenchen~Li}
\email{lcc@lnnu.edu.cn}
\affiliation{%
  \institution{Liaoning Normal University}
   \country{China}
}

\author{Shenglin Wang}
\email{swang586-c@my.cityu.edu.hk}
\affiliation{%
  \institution{City University of Hong Kong, Hong Kong Special Administrative Region of China}
   \country{China}
}


\author{Furui Zhan}
\email{izfree@dlmu.edu.cn}
\affiliation{%
  \institution{Dalian Maritime University}
   \country{China}
}


\renewcommand{\shortauthors}{Ji et al.}


\begin{abstract}

With the continuous growth in the scale and complexity of software systems, defect remediation has become increasingly difficult and costly. 
Automated defect prediction tools can proactively identify software changes prone to defects within software projects, thereby enhancing software development efficiency. 
However, existing work in heterogeneous and complex software projects continues to face challenges, such as struggling with heterogeneous commit structures and ignoring cross-line dependencies in code changes, which ultimately reduce the accuracy of defect identification.
To address these challenges, we propose an approach called RC\_Detector, which learns hidden semantic representations of code lines by incorporating dependencies between them to detect the root causes in bug-fixing commits. 
RC\_Detector comprises three main components: the bug-fixing graph construction component, the code semantic aggregation component, and the cross-line semantic retention component. 
The bug-fixing graph construction component identifies the code syntax structures and program dependencies within bug-fixing commits and transforms them into heterogeneous graph formats by converting the source code into vector representations. 
The code semantic aggregation component adapts to heterogeneous data by using heterogeneous attention to learn the hidden semantic representation of target code lines.
The cross-line semantic retention component regulates propagated semantic information by using attenuation and reinforcement gates derived from old and new code semantic representations, effectively preserving cross-line semantic relationships.
Extensive experiments were conducted to evaluate the performance of our model by collecting data from 87 open-source projects, including 675 bug-fixing commits. The experimental results demonstrate that our model outperforms state-of-the-art approaches, achieving significant improvements of 83.15\%,96.83\%,78.71\%,74.15\%,54.14\%,91.66\%,91.66\%, and 34.82\% in MFR, respectively, compared with the state-of-the-art approaches.

\end{abstract}



\maketitle
\section{INTRODUCTION}
After the software is deployed, hidden bugs are inevitably exposed under certain operating environments, necessitating continuous maintenance and corrections by developers to extend the software's life cycle. 
The process of identifying and fixing software defects often requires substantial effort, consuming significant time, money, and human resources~\cite{xia2016predicting}. 
Reducing the cost associated with defect identification and remediation has been a major focus for software developers. 
Just-in-time (JIT) defect prediction techniques~\cite{1,2,24}  have always attracted considerable attention in the field of software engineering
Unlike traditional defect prediction approaches that work on entities with coarse-grained features (such as files, modules, or packages~\cite{xia2016predicting,lessmann2008benchmarking,hassan2009predicting,nam2015heterogeneous,xia2016hydra}), JIT defect prediction focuses on change-level predictions at the commit level. This fine-grained approach allows for immediate predictions as new code changes occur, enabling developers to address potential defects in real-time rather than during broader code review or testing phases. Additionally, modern JIT defect prediction techniques emphasize the use of end-to-end deep learning frameworks~\cite{7}, which mitigate the reliance on handcrafted features and statistical techniques commonly seen in traditional approaches.

In JIT defect prediction, the SZZ algorithm~\cite{3} plays a crucial role. 
The SZZ algorithm is a change analysis approach used in software engineering that significantly reduces developers' workload by accurately locating bug-inducing changes. 
It is primarily employed to identify and track bug-inducing changes in software projects. 
By tracing the change history in version control systems, the SZZ algorithm helps developers pinpoint which code changes led to software defects, thereby improving software reliability and maintainability.

The original SZZ algorithm (B-SZZ) was proposed by Sliwerski et al.~\cite{3} and designed to trace the last changes made to lines deleted or modified in bug-fixing commits and label those changes as bug-inducing commits. 
However, B-SZZ's implementation is limited by its simplistic tracking of code line changes, making it prone to misjudgment when encountering noise. 

To improve the accuracy of this algorithm, researchers have proposed various modifications that improve B-SZZ in different dimensions.
To address the noise issue in B-SZZ, Kim et al.~\cite{4} introduced AG-SZZ, which filters out blank lines, comment lines, and cosmetic changes in the code using an annotation graph. This enhancement allows AG-SZZ to more accurately identify true bug-inducing code commits, significantly improving the algorithm's precision. 
Da Costa et al.~\cite{5} extended the SZZ algorithm by filtering out meta changes (such as branches, merges, and property changes), proposing MA-SZZ. 
Since these meta changes do not genuinely modify the source code, MA-SZZ reduces the likelihood of false positives by excluding these invalid code modifications. However, as the complexity of codebases increases, refactoring operations pose a significant challenge for the SZZ algorithm. 
To address this, Neto et al.~\cite{6} proposed RA-SZZ, which integrates refactoring detection tools, RefDiff and RefactoringMiner, to reduce false positives by identifying and excluding refactoring operations. 
This further optimizes RA-SZZ's performance in complex codebases. 

Despite the progress made by existing work, one challenge remains: In software engineering projects, bug-fixing commits often include many non-essential changes~\cite{21,22}, which can be highly heterogeneous and complex. Essential bug-fixing part coexists with other forms of modification, such as code refactoring, feature additions, etc. Traditional approaches either oversimplify these heterogeneous changes (assuming that all changed lines have repair characteristics: original SZZ~\cite{3}) or separate them according to rigid predefined classification schemas for different code types or categories of changes (RA-SZZ~\cite{6}). However, both strategies rely on static rules to solve the problem, making it difficult to pinpoint the difference between the actual fixed lines and other lines that have been changed for refactoring or enhancement.

Neural SZZ was proposed by Tang et al.~\cite{tang2023neural}to solve this problem, Neural SZZ is a deep learning-based approach that captures the semantic relationships between deleted lines and other logically related lines by constructing heterogeneous graphs of commits and using a heterogeneous graph attention network(HAN) model. 
This algorithm evaluates and sorts based on the semantics of these deleted lines, offering a more intelligent and accurate approach for identifying the root causes in bug-fixing commits.

Using Semantic associations between adjacent or logically related lines are critical to understanding code changes, the functional logic of code is often scattered across multiple lines or blocks of code~\cite{23,shao2008understanding,yin2024line}. Deep learning architectures often attempt to understand the "more global" representation of lines of code through multiple layers of aggregation. However, with the increase of the range of code line interaction information, the local semantic information learned early is overshadowed by the global information, resulting in the loss of key semantic features. It is difficult to effectively capture and preserve cross-line semantic relationships, which will make the semantics of each code line tend to be homogenized ~\cite{jin2022feature,li2018deeper}, and its own personality will disappear.

To address the challenge, we propose a new heterogeneous graph neural network model called RC\_Detector to learn the hidden semantic representations of code lines. 
RC\_Detector consists of three main components: the bug-fixing graph construction component, the code semantic aggregation component, and the cross-line semantic retention component.
Specifically, the bug-fixing graph construction component extracts the source code from both the previous and newer versions, generating the corresponding syntax trees and program dependency graphs to identify the code syntax structures and program dependencies within the bug-fixing commits. 
The code lines are then mapped to graph nodes based on this information, and the types of edges between nodes are determined.  
Finally, the source code is converted into vector representations, forming a heterogeneous graph format suitable for neural networks.
The code semantic aggregation component processes each heterogeneous graph generated from bug-fixing commits by calculating the semantic similarity between the target code line node and related code line nodes, assigning weights to all related code lines accordingly. 
These weights are then used to aggregate the semantics of code lines directly related to the target, thereby learning the hidden semantic representation of the target code line. 
The cross-line semantic retention component calculates the attenuation and reinforcement gates using the semantic representations of the old and new code.
These gates dynamically manage the flow of information by controlling the retention or updating of the original old code semantics and the aggregated new semantic representations to preserve cross-line semantic relationships. 

The bug-fixing graph construction and code semantic aggregation components draw on the existing heterogeneous graph neural network technology ~\cite{tang2023neural}, and we improve the code semantic aggregation components. Our main contribution is the introduction of the cross-line semantic retention component, which addresses the deficiency of the existing model in preserving cross-line semantic relations during bug fixes.

To evaluate the effectiveness of our model, we conducted experiments using data from 87 open-source projects, comprising a total of 675 bug-fixing commits, and compared the results with state-of-the-art approaches.  
Since developers often need to quickly identify and address the most critical issues, we assessed RC\_Detector's performance using the Recall@N metric, with \emph{N} set to 1, 2, and 3, and estimated the model's cost-effectiveness using the mean first rank (MFR).
Our model achieved improvements of 4.32\%, 7.06\%, 4.81\%, and 34.82\% over the best state-of-the-art approach in recall@1, recall@2, recall@3, and MFR, respectively.  
These experimental results demonstrate the effectiveness of using RC\_Detector to capture the semantic relationships between each deleted line and other deleted or added lines.

In summary, the main contributions of this work can be summarized as follows:
\begin{itemize}
\item Based on semantic aggregation~\cite{tang2023neural}, we propose the RC\_Detector method to address the limitations of ignoring cross-line dependencies in the previous method. This method preserves key early local information through a gating mechanism, reduces the homogeneity of the semantic representation of code lines, and improves the defect prediction performance of JIT methods.
\item We conducted extensive experiments on a dataset comprising 675 bug-fixing commits from 87 open-source projects to evaluate the impact of RC\_Detector on defect prediction performance. The experimental results demonstrate that our model outperforms state-of-the-art approaches,
achieving a 34.82\% improvement in MFR compared to the best state-of-the-art approaches.
\item We have made our code and experimental dataset publicly available as open-source, which
can benefit the research community and foster further development in this field~\cite{myreplication}.
\end{itemize}

The remainder of this paper is organized as follows: 
Section 2 discusses related work, 
Section 3 introduces the main components of RC\_Detector. Experimental settings and results are presented in Sections 4 and 5, respectively. 
Section 6 elaborates on threats to validity. 
Finally, Section 7 concludes this work and outlines potential future directions.

\section{RELATED WORKS}

In this section, we explore several studies closely related to Just-In-Time (JIT) defect prediction, which form the foundation of our work.

Just-In-Time (JIT) defect prediction aims to predict potential defects introduced during code commits in a timely manner, thereby helping developers identify and fix potential issues at an early stage. 
Early JIT defect prediction approaches primarily relied on traditional machine learning approaches, encompassing several steps such as feature extraction, data labeling, and model construction~\cite{kim2008classifying}. 
Feature extraction involved manually extracting various attributes from software version control systems to describe code changes. 
Data labeling typically uses the SZZ algorithm to trace and label commits that introduced defects. 
The constructed models were then trained on labeled data and features to predict whether unlabeled commits might introduce defects.

In recent years, JIT defect prediction has transcended the limitations of traditional approaches by incorporating more sophisticated and innovative techniques. 
For example, Huang et al. introduced~\cite{7} an end-to-end deep learning framework that uses Convolutional Neural Networks (CNN) to automatically generate features from commit messages and code changes, followed by a fully connected layer for defect prediction. 
Building on this, Choi et al. proposed CC2Vec~\cite{8}, which leverages a Hierarchical Attention Network (HAN) to automatically learn distributed representations of code commits, improving performance. 
To further enhance the efficiency of JIT defect prediction, Hoang et al. developed JITLine~\cite{9}, an approach that combines the strengths of DeepJIT and CC2Vec, further improving prediction accuracy and granularity. 
Neural SZZ, proposed by Tang et al.~\cite{tang2023neural}, extends the traditional SZZ algorithm by employing Heterogeneous Graph Neural Networks (HGNN) to capture deep semantic representations of code, focusing on identifying the root causes of defects through learning-to-rank techniques. Neural SZZ does not rely solely on superficial textual differences or syntactic rules, but adds processing related to semantic features. First, syntactic features are obtained using an abstract syntax tree. Second, a vectorised representation of the source code is obtained, and a deep learning model is used to capture semantic features. Finally, a ranking algorithm is used to recommend the line that is the root cause of the bug.

In addition to adopting novel defect prediction approaches, researchers have also explored data augmentation to enhance defect prediction capabilities~\cite{xu2024code,1,10571907}. 
Kamei et al. proposed data augmentation techniques~\cite{1} that synthesize additional training samples or transform existing samples to alleviate the problem of data sparsity. 
These approaches significantly improve model robustness without incurring additional data collection costs. 
Moreover, Tsuda et al. utilized anomaly detection techniques~\cite{10571907} (such as isolation forest) to reduce noise by identifying and filtering out potentially mislabeled samples, thereby improving the model's predictive accuracy.

\section{RC\_Detector MODEL} \label{MODEL}

In this section, we provide a detailed description of the entire model algorithm. 
Section 3.1 gives an overview of the RC\_Detector model framework and the flow of the algorithm. 
Then, in Sections 3.2 to 3.5, we delve into the specific components of RC\_Detector.

\begin{figure*}
	\centering
	\includegraphics[width=\linewidth]{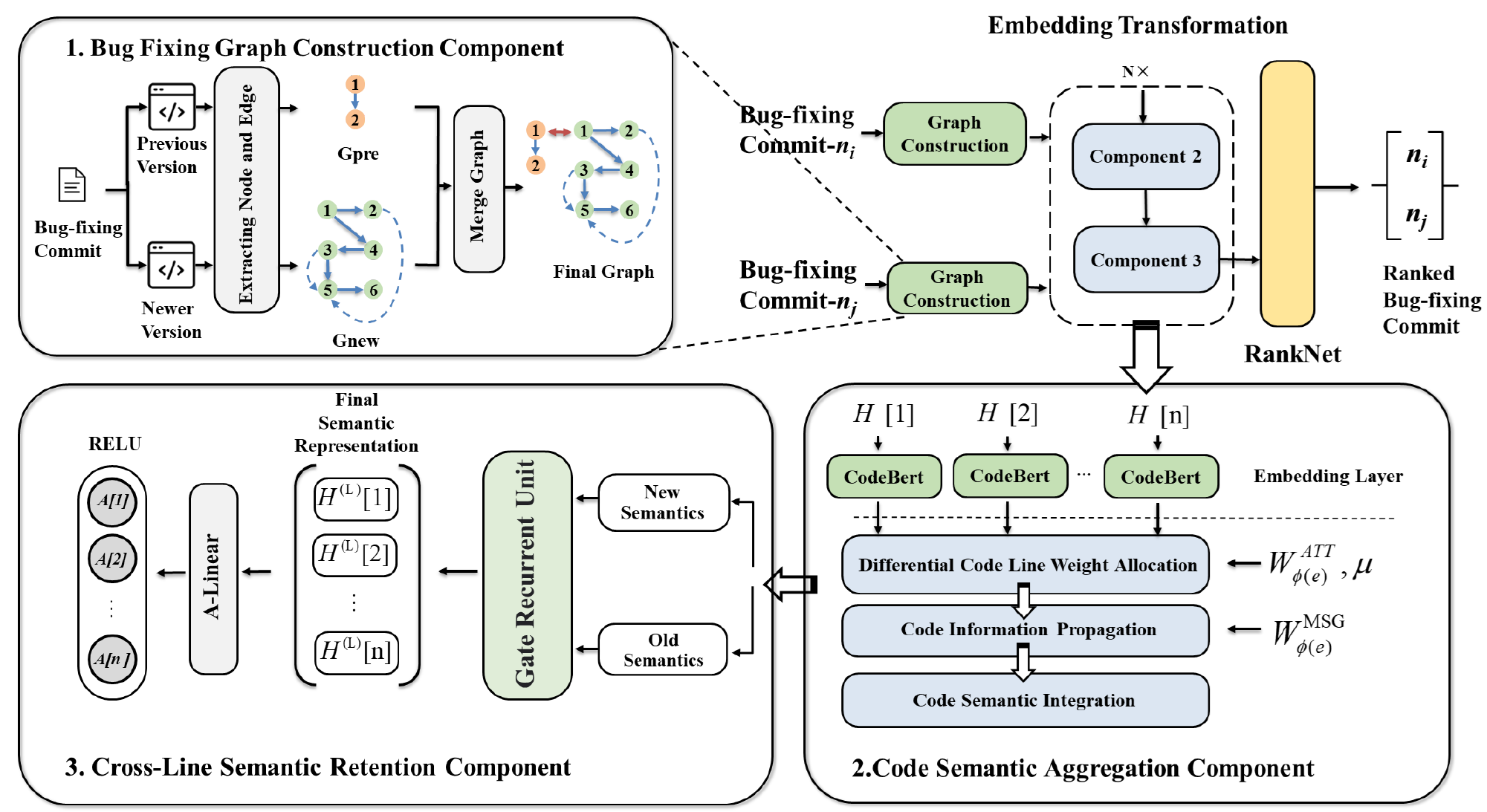}
	\caption{The overall framework of RC\_Detector}
	\label{fig:framework}
\end{figure*}

\subsection{Overview}\label{AA}

To accommodate the heterogeneity of code commits while accurately capturing the semantics of contextual code lines~\cite{25}, thereby improving the defect prediction performance of JIT approaches in projects, we propose the RC\_Detector model. The framework of the RC\_Detector model is illustrated in  Fig~\ref{fig:framework}. RC\_Detector primarily consists of three components: the bug-fixing graph construction component, the code semantic aggregation component, and the cross-line semantic retention component.

The bug-fixing graph construction component generates corresponding syntax trees and program dependency graphs by extracting source code from both the previous and newer versions, enabling the identification of code syntax structures and program dependencies in bug-fixing commits.
Based on this information, code lines are mapped to graph nodes, and the types of edges between nodes are determined.
The source code is then transformed into vector representations, creating a heterogeneous graph format suitable for neural networks. 
These graphs are subsequently input in batches into the code semantic aggregation component.

The code semantic aggregation component assigns attention weights to all relevant code lines within a heterogeneous graph, which is constructed from bug-fixing commits. 
This assignment is based on the semantic similarity between the target code line nodes and the relevant code line nodes~\cite{vaswani2017attention}, as well as the dependencies between them. 
The component then integrates the semantic information from various program dependencies according to these weights, allowing it to learn the hidden semantic representation of the target code line.

The cross-line semantic retention component acquires attenuation gates and reinforcement gates~\cite{chung2014empirical} based on the old and new semantic representations of the code. 
The attenuation gate determines which information needs to be discarded, while the reinforcement gate determines which new information needs to be added to the current node representation. This regulating of transitive semantic information helps capture semantic dependencies across code lines, ensuring that crucial cross-line relationships are effectively preserved.

After processing through the RC\_Detector model, the final semantic vector is input into the RankNet model~\cite{burges2010ranknet} to learn the relative priority of the deletion nodes. 
RankNet assigns a score to each bug-fixing commit based on the input semantic vectors, ranking the bug-fixing commits to place the ones most likely to introduce bugs at a higher rank, thereby identifying the root cause of the bugs.

\subsection{Bug-Fixing Graph Construction Component} 

In this section, the component constructs a heterogeneous graph format suitable for neural networks by analyzing bug-fixing commits~\cite{tang2023neural}. Specifically, this process involves two sub-steps:

\subsubsection{Graph Construction}
\
\par For a bug-fixing commit, the first step is to extract the Java source code from the previous and newer versions. Then, use the JavaParser tool~\cite{javaparser}  to construct their respective Abstract Syntax Trees (ASTs), referred to as ASTpre and ASTnew.  
Subsequently, the static analysis tool Joern~\cite{yamaguchi2014modeling} is employed to construct the Program Dependency Graphs (PDGs).
The PDGs includes the Control Flow Graph (CFG)~\cite{allen1970control}, Data Dependency Graph (DDG)~\cite{ferrante1987program} , Call Graph (CG)~\cite{ryder1979constructing}, and Class Member Reference Graph (CMFG). 
Control flow or data flow edges are then added between corresponding nodes in the two versions of the PDGs using a depth-first search algorithm, generating the previous version's graph (Gpre) and the newer version's graph (Gnew).  
Finally, a line-mapping algorithm in conjunction with the GumTree tool~\cite{falleri2014fine}is used to merge these two graphs, connecting matched deletion and addition nodes with a line mapping edge to obtain the final heterogeneous graph. 
The heterogeneous graph is defined as 
~\cite{hu2020heterogeneous}: ${\rm{G}} = (V,E,A,R)$,
where each code line corresponds to a node $v$ satisfying $\forall v \in V$, each program dependency relation $e$ satisfying $\forall e \in E$ and the code line type mapping function $\tau(v): V \to A$ and the program dependency relation mapping function $\phi(e): E \to R$ are defined accordingly.

\subsubsection{Node Embedding Transformation}
\
\par For the heterogeneous graph constructed in the previous step, each node corresponds to a line of code (e.g., a deleted or added line).
First, CodeBERT~\cite{feng2020codebert} is employed as the node embedding layer to convert the source code of each line into a fixed-length vector representation. 
CodeBERT is a widely adopted pre-trained language model that has demonstrated superior performance across various code-related tasks~\cite{feng2020codebert}.
It effectively captures the semantics of code statements, providing rich node representations suitable for graph neural networks. 
For each code node $i$, we use CodeBERT to obtain its corresponding initial semantic representation ${H^0}[i]$.

The constructed heterogeneous graph is then divided into batches of equal size and input into the code semantic aggregation component for the aggregation of the semantic vector representation of code lines.

\subsection{Code Semantic Aggregation Component}  

We conduct code semantic aggregation based on Heterogeneous Graph Transformer (HGT)~\cite{hu2020heterogeneous}. Heterogeneous Graph Transformer (HGT) is a graph neural network architecture designed specifically for heterogeneous graphs (containing multiple types of nodes and edges). In our context, the node types are lines that are deleted or added, and the edge types are control flow edges, data flow edges, and call edges. HGT dynamically computes the attention weights between nodes through a multi-head mechanism, weighted aggregates semantic information from neighbors, and updates the information of the target node.
Specifically, the code semantic aggregation component is composed of three parts: (1) Differential Code Line Weight Allocation,(2) Code Information Propagation, and (3) Code Semantic Integration.

\subsubsection{Differential Code Line Weight Allocation}
\
\par During the code-fixing process, different types of code lines (deleted lines and added lines) may exhibit significant semantic and structural differences. 
To better capture these differences and avoid information conflation, the model defines a separate set of projection matrices for deleted and added lines. 
These matrices map the code lines into feature representations in distinct ways, enabling the model to accurately reflect the characteristics of each type of code line.

The semantic representation of each code line node ${H^{(l - 1)}}[{\rm{i}}]$, obtained in the previous stage, is mapped into Key, Query, and Value vectors ($K{^{}}({\rm{i}})$,$Q{^{}}({\rm{i}})$,$V{^{}}({\rm{i}})$) for use in subsequent attention 
calculations~\cite{vaswani2017attention}:

\begin{equation}
{K^{}}(i) = K\text{-}Linear_{\tau (i)}^{}\left( {H^{(l - 1)}[i]} \right)\forall i \in V
\end{equation}

\begin{equation}
{Q^{}}(i) = Q\text{-}Linear_{\tau (i)}^{}\left( {H^{(l - 1)}[i]} \right)\forall i \in V
\end{equation}

\begin{equation}
{V^{}}(i) = V\text{-}Linear_{\tau (i)}^{}\left( {H^{(l - 1)}[i]} \right)\forall i \in V
\end{equation}
where $\tau (i)$ represents the type of the code line $i$  and  $K - Linear,Q - Linear,V - Linear$ represent linear projection.

To more finely characterize the complex associations between code lines, the Key, Query, and Value vectors are split into multiple attention heads~\cite{vaswani2017attention}. ${R^D} \to {R^{\frac{D}{H}}}$,
where $H$ is the number of attention heads, and $\frac{D}{H}$ is the dimensionality of each head. Specifically:

\begin{equation}
    K(i) = \text{Concat}([{K^1}(i), {K^2}(i), \ldots, {K^H}(i)])
\end{equation}

\begin{equation}
    Q(i) = \text{Concat}([{Q^1}(i), {Q^2}(i), \ldots, {Q^H}(i)])
\end{equation}

\begin{equation}
    V(i) = \text{Concat}([{V^1}(i), {V^2}(i), \ldots, {V^H}(i)])
\end{equation}
where ${K^{1 \sim {\rm{H}}}}(i)$ represents the attention heads of the Key vector of code line $i$.
Similarly, this applies to the Query and Value vectors.

By independently calculating attention in each head, the model can better understand the complex interactions within the code, ultimately enhancing its performance in capturing subtle patterns and dependencies.

In bug-fixing commits, we often need to further analyze the dependency relationships between source and target code lines. For a given code line pair ${\rm{e = }}(s,t)$
, RC\_Detector represent their relationship through the triple $ < \tau (s),\phi (e),\tau (t) > $
, where $\tau (s)$ and $\tau (t)$ represent the types of the source code line and the target code line, respectively, and $\phi (e)$ describes the specific dependency relationship between them, such as control dependency or data dependency.

Next, the component calculates the semantic similarity between the source code line and the target code line. HGT computes the similarity between ${K^i}(s)$ and ${Q^i}(t)$, where ${K^i}(s)$ is the Key vector of the $i$-th head of the source code line s, and ${Q^i}(t)$ is the Query vector of the $i$-th head of the target code line. 
To enable the model to differentiate between diverse code line relationships, HGT assigns a distinct learnable matrix 
$W_{\phi (e)}^{ATT} \in {R^{\frac{D}{{\rm{H}}}}} \times {R^{\frac{D}{{\rm{H}}}}}$ for each type of dependency relationship $\phi (e)$. 
In this way, even the same pair of code lines can learn different semantic information through different program dependencies. 
Each dependency relationship is associated with a learnable prior tensor 
$\mu \langle \tau (s),\phi (e),\tau (t)\rangle  \in {{\rm{R}}^{\left| {\rm{A}} \right| \times \left| R \right| \times \left| A \right|}}$, which can adjust the attention weights according to the specific dependency type. 
The attention weights for each attention head are then given by the formula:
 
\begin{equation}
ATT\text{-}hea{d^i}(s,e,t) = \left( {K^i(s)W_{\phi (e)}^{ATT}Q^i{(t)^T}} \right) \cdot \frac{\mu \langle \tau (s),\phi (e),\tau (t)\rangle}{\sqrt{d}}
\end{equation}
where the dimension $d$ of linear projection $K - Linear$ and $Q - Linear$  is used as a scaling factor to ensure that the inner product  ${{K^i}(s)W_{\phi (e)}^{ATT}{Q^i}{{(t)}^T}} $does not become excessively large, thereby maintaining the stable operation of the Softmax function.

For each target code line node $t$, the attention weights with all its corresponding source code line nodes are calculated. 
The attention heads $ATT - hea{d^i}(s,e,t)$ are concatenated into ${\rm{Concat - Head}}(s,e,t) \in {R^{H \times {\rm{1}}}}$ and then passed through a softmax function to normalize the attention weights across all source code line nodes on each head, ensuring that the sum equals $1$, yielding the final attention weights:

\begin{equation}
Attentio{n_{HGT(s,e,t)}} = \mathop {S{\rm{oftmax}}}\limits_{\forall s \in N(t)} {\rm{(Concat - Head}}(s,e,t))
\end{equation}

\begin{equation}
{\rm{Concat - Head}}(s,e,t) = \mathop {||}\limits_{i \in [1,h]} ATT - hea{d^i}(s,e,t)
\end{equation}

Specifically:

\begin{equation}
\sum\limits_{\forall s \in N(t)} Attentio{n_{HGT(s,e,t)}}  = 1_{H \times {1}}
\end{equation}

\begin{equation}
Attentio{n_{HGT(s,e,t)}} = [{p_s}^1,{p_s}^2, \ldots ,{p_s}^H]
\end{equation}

\begin{equation}
{p_s}^i = \frac{{esp({a_s}^i)}}{{\sum\limits_{\forall s \in N(t)} {exp({a_s}^i)} }}
\end{equation}

\begin{equation}
{a_s}^i = ATT - hea{d^i}(s,e,t)
\end{equation}
where $\mathop {||} $ is the same as $\text{Concat}$ , ${a_s}^i$ represents the attention weights of the $i$-th head of source line $s$, and ${Attentio{n_{HGT(s,e,t)}}}$ represents the multiple attention of source code line $s$ for the target code line $t$.

\subsubsection{Code Information Propagation}
\
\par In the code repair task, dependencies between different code lines can manifest in various forms, such as control and data dependencies, among others.  
To comprehensively capture these diverse dependencies, the model must be capable of flexibly transmitting different types of information between code lines based on the specific dependency types.

Specifically, during the process of propagating code information, the model defines an information transmission matrix $W_{\phi (e)}^{MSG} \in {R^{\frac{D}{{\rm{H}}}}} \times {R^{\frac{D}{{\rm{H}}}}}$ for each type of dependency relationship. 
This trainable matrix adjusts the manner in which information is transmitted between code lines, ensuring that the characteristics of each dependency are accurately captured and expressed.
The multi-head Value vectors of each code line obtained earlier are multiplied by their transmission matrix to generate their respective heterogeneous information heads:

\begin{equation}
MSG\text{-}hea{d^i}(s,e,t) = V^i(s)W_{\phi (e)}^{MSG}
\end{equation}
where ${V^i}(s)$ represents the content of the $i$-th head of the Value vector corresponding to source code line $s$.

The different message heads are then concatenated to form the heterogeneous message representation for the target node:

\begin{equation}
Messag{e_{HGT}}(s,e,t) = \mathop {||}\limits_{i \in [1,h]} MSG - hea{d^i}(s,e,t)
\end{equation}

Through this approach, we effectively combine program dependency relationships with the semantic representation of the code itself, resulting in a richer and more informative representation.  
This enriched representation is then utilized in the subsequent Code Semantic Integration phase, providing a solid foundation for accurately modeling the complexities of code changes.

\subsubsection{Code Semantic Integration}
\
\par After completing code line weight allocation and code information propagation, we need to aggregate the different types of information from the source nodes to update the semantic representation of the target nodes. 
This process is achieved by performing a weighted sum of the previously calculated attention weights $Attentio{n_{HGT(s,e,t)}}$ and the heterogeneous information $Messag{e_{HGT}}(s,e,t)$ from the source nodes to obtain the updated neighbor aggregation vector:

\begin{equation}
\mathop{H}\limits^{\sim}{}^{(l)}[t] = \mathop{\bigoplus}\limits_{\forall s \in N(t)} \left( \text{Attention}_{HGT(s,e,t)} \cdot \text{Message}_{HGT(s,e,t)} \right)
\end{equation}
Specifically, the calculation is as follows:

\begin{equation}
\mathop{H}\limits^{\sim}{}^{(l)}[t] = [{W_t}^1, \ldots, {W_t}^H]
\end{equation}

\begin{equation}
{W_t}^i = \sum\limits_{\forall s \in N(t)} ATT\text{-}hea{d^i}(s,e,t) \times MSG\text{-}hea{d^i}(s,e,t)
\end{equation}
where $\mathop{\bigoplus}$ represents element-wise addition,and $\forall s \in N(t)$ indicates all source nodes $s$ of the target node $t$. 

In this step, the model aggregates semantic information from various program dependencies using attention weights.  
By applying these attention-weighted influences, the model captures the most critical dependencies and relationships necessary for understanding the semantics of the current code line.  
The obtained contextual code semantic information is then passed to the cross-line semantic retention component for further processing.

\subsection{Cross-Line Semantic Retention Component}  

\begin{table}[!t]
\centering
\caption{Notation used in Gate Mechanism}
\label{tab:gate_mechanism}
\begin{adjustbox}{max width=\textwidth}
\begin{tabular}{>{\centering\arraybackslash}m{2.5cm} >{\raggedright\arraybackslash}m{10cm}}
\toprule
\textbf{Notation} & \textbf{Definition} \\ 
\midrule
$W_{ir}$ & Weight matrix for inputs to the \textbf{attenuation gate} \\ 
$b_{ir}$ & Bias vector for inputs to the \textbf{attenuation gate} \\ 
$W_{hr}$ & Weight matrix from historical semantic representation to the \textbf{attenuation gate} \\ 
$b_{hr}$ & Bias vector from historical semantic representation to the \textbf{attenuation gate} \\ 
$W_{in}$ & Weight matrix for inputs to the candidate hidden state \\ 
$b_{in}$ & Bias vector for inputs to the candidate hidden state \\ 
$W_{hn}$ & Weight matrix from historical semantic representation to the candidate hidden state \\ 
$b_{hn}$ & Bias vector from historical semantic representation to the candidate hidden state \\ 
$W_{iz}$ & Weight matrix for inputs to the \textbf{reinforcement gate} \\ 
$b_{iz}$ & Bias vector for inputs to the \textbf{reinforcement gate} \\ 
$W_{hz}$ & Weight matrix from historical semantic representation to the \textbf{reinforcement gate} \\ 
$b_{hz}$ & Bias vector from historical semantic representation to the \textbf{reinforcement gate} \\ 
$\sigma$ & Sigmoid activation function \\ 
$tanh$ & Tanh activation function \\ 
$\odot$ & Element-wise multiplication\\
\bottomrule
\end{tabular}
\end{adjustbox}
\end{table}

To effectively capture and preserve cross-line semantic relationships, RC\_Detector introduces a controlled Gated Recurrent Unit (GRU)~\cite{chung2014empirical}.

\subsubsection{Gate Mechanism}
\
\par The GRU controls the update of hidden states through attenuation gates and reinforcement gates, effectively regulating the transmitted semantic information. For the current processing stage $l$, RC\_Detector uses the GRU to calculate the final semantic representation $H^{(l)}[t]$ of the node $t$ in the current processing step. 
The gate mechanism acquires two gating states, attenuation gate and reinforcement gate, by using the historical semantic representation $H^{(l - 1)}[t]$ and the currently obtained neighbor aggregation vector$\mathop{H}\limits^{\sim}{}^{(l)}[t]$. 
The meanings of all the parameters used are summarized in Table \ref{tab:gate_mechanism}.

\textbf{Attenuation Gate.}The main function of attenuation gates is to determine to what extent the current semantic information depends on the results of previous processing $H{}^{(l - 1)}[t]$. 
The computations involved are represented as follows:

\begin{equation}
r = \sigma \left( {W_{ir}} \mathop{H}\limits^{\sim}{}^{(l)}[t] + {b_{ir}} + {W_{hr}} H^{(l - 1)}[t] + {b_{hr}} \right)
\end{equation}

\textbf{Reinforcement Gate.} The reinforcement gate determines the extent to which the historical semantic representation $H{}^{(l - 1)}[t]$ is preserved for the current processing step, with parameters calculated as follows:

\begin{equation}
z = \sigma \left({W_{iz}}\mathop{H}\limits^{\sim}{}^{(l)}[t] + {b_{iz}} + {W_{hz}}H^{(l - 1)}[t] + {b_{hz}} \right)
\end{equation}

\textbf{Candidate Semantic Representation.} Based on the neighbor aggregation vector $\mathop{H}\limits^{\sim}{}^{(l)}[t]$ and the historical semantic representation$H{}^{(l - 1)}[t]$ adjusted by the attenuation gate, the candidate semantic representation for the current processing step is calculated accordingly:

\begin{equation}
n = \tanh \left( {W_{in}} \mathop{H}\limits^{\sim}{}^{(l)}[t] + {b_{in}} + r \odot \left( {W_{hn}} H^{(l - 1)}[t] + {b_{hn}} \right) \right)
\end{equation}

Finally, the reinforcement gate is used to perform a weighted combination of the historical semantic representation and the candidate semantic representation for the current processing step, yielding the final semantic representation of the target code line node at the current processing step:

\begin{equation}
{H^{(l)}}[t] = (1 - z) \odot n + z \odot {H^{(l - 1)}}[t]
\end{equation}

The main purpose of using the Gated Recurrent Unit is to enhance the model's ability to capture and retain key semantic features across code lines. The GRU accomplishes this by adopting gating mechanisms, specifically attenuation and reinforcement gates, to effectively manage the flow of information.

\subsubsection{Task-Specific Mapping Mechanism}
\
\par The task-specific mapping mechanism aims to map the semantic representation vector of the target node, processed through multiple layers of the network, to a space suitable for the current task (e.g., classification, ranking, prediction). 
This step is intended to adapt the node vector output by the model to the requirements of the specific task, thereby improving the quality of task completion. 
For this purpose,  RC\_Detector applies a linear projection to the final semantic representation of the code and uses the ReLU activation function to introduce nonlinearity, allowing for the fitting of more complex relationships.Task embedding representation $A[t]$ is calculated as follows:

\begin{equation}
A[t] = {\mathop{\rm Re}\nolimits} lu({{\rm{W}}_{proj}}{H^{(l)}}[t] + {b_{proj}})
\end{equation}

\begin{equation}
ReLU(x) = max(0,x)
\end{equation}
where ${{\rm{W}}_{proj}}$ denotes the weight matrix in the linear transformation, ${b_{proj}}$ denotes the bias vector in the linear transformation

\subsection{Pairwise Ranking of Bug-Fixing Commits} 
After obtaining the downstream task embedding representation of each node, we further train the model to rank bug-fixing commits.
Specifically, RC\_Detector utilizes the RankNet model same to NeuralSZZ ~\cite{tang2023neural}. RankNet is a pairwise ranking approach that has proven effective in real-world ranking problems~\cite{karmaker2017application,chapelle2011yahoo,song2014adapting}. 
The RankNet model is trained by learning the relative priority of deleted nodes in a pairwise manner. 
For each bug-fixing commit, we pair it with other bug-fixing commits to obtain a series of pairs $< n_{i},n_{j} >$. 
RankNet processes one of the bug-fixing commit pairs at a time and obtains the corresponding scores based on the task embedding representations of the deletion nodes in each commit, denoted as $s_{i}$ and $s_{j}$. 
The probability that bug-fixing commit $n_{i}$ ranks higher than $n_{j}$ is then calculated accordingly:

\begin{equation}
{P_{\emph{ij}}} = \frac{1}{{1 + {e^{ - \sigma (s_{i} - s_{j})}}}}
\end{equation}
The ground truth probability for the relative priority of bug-fixing commits is:

\begin{equation}
\overline{P}_{\emph{ij}} = 
\begin{cases} 
1 & \text{if } n_{i} \text{ is root cause node and } n_{j} \text{ is not} \\
0 & \text{if } n_{j} \text{ is root cause node and } n_{i} \text{ is not} \\
0.5 & \text{otherwise}
\end{cases}
\end{equation}
Finally, the RankNet model is trained using the cross-entropy loss function:

\begin{equation}
L = -\overline{P_{ij}} \log(P_{ij}) - (1 - \overline{P_{ij}})\log(1 - P_{ij})
\end{equation}

By training RankNet, bug-inducing commits are effectively separated from other commits and ranked higher, enabling practitioners to promptly identify the root causes of bugs within code commits.

\subsection{Comparison with NeuralSZZ:}
Inspired by the NeuralSZZ method, RC\_Detector also uses a framework that combines syntactic structure analysis with semantic representation learning to process bug-fixing commits. By fusing the syntactic features of the code with the semantic features of deep learning, as well as mature deep learning techniques, it effectively improves the accuracy of identifying bug-inducing commits.

NeuralSZZ uses a heterogeneous attention network (HAN) to capture the semantic relationships between lines of code. A heterogeneous attention network (HAN) generally consists of two levels: ‘node-level attention’ and ‘semantic-level attention.’ Node-level attention is used to learn the influence weights between the target node and its context nodes in the same metpath neighborhood; semantic-level attention re-weights the aggregation of multiple metpaths to extract the most useful semantic information for the final task. However, attention is usually calculated at the graph level or the meta-path level, and the meta-paths need to be defined first, which requires additional engineering and algorithm design.

RC\_Detector constructs a hybrid neural network architecture based on the Heterogeneous Graph Transformer (HGT) and gated recurrent unit (GRU) based on modeling heterogeneous graph structures. HGT is a heterogeneous graph attention network that models meta-relations and only takes single-hop edges as input. It projects different node(edge) types into a unified semantic space by inducing a type-parameterised mapping mechanism, and performs semantic aggregation by emulating a transformer-style attention mechanism. It stacks multiple network structures to pass on information from higher-order neighbours of different types, thereby implicitly learning and extracting ‘meta-paths’ that are more important for different downstream tasks. In other words, the semantics of the code line nodes contained in these implicit meta-paths are more important for downstream tasks. At the same time, the gating mechanism of the GRU is used to retain and update local contextual information during the gradual expansion of the receptive field, preventing the local semantic information learned early from being overshadowed by global information and retaining the individual semantics of each code line.

\section{EXPERIMENTAL SETUP} \label{Experimental}

\subsection{Datasets}
We conducted our experimental study using a comprehensive dataset ~\cite{tang2023neural}composed of three sub-datasets containing bug-fixing and bug-inducing commits. These datasets are of higher quality and contain less noise compared to those generated by the SZZ algorithm.

\textbf{DATASET1.} Collected by Wen et al.~\cite{wen2019exploring}, this dataset was manually reviewed by the authors across various projects and supplemented with automated tools to ensure the accuracy and completeness of the data.

\textbf{DATASET2.} Built by Song et al.~\cite{song2022regminer}, this dataset identifies bug-inducing and bug-fixing commits by utilizing existing test cases in the codebase. A commit is considered bug-inducing if it causes a previously passing test case to fail, and a subsequent commit that makes the test case pass again is considered a bug-fixing commit.

\textbf{DATASET3.} Collected by Neto et al.~\cite{neto2019revisiting}, this dataset leverages detailed information from the Defects4J dataset~\cite{just2014defects4j}, including change logs and patch files from version control systems. By combining manual and automated approaches, they meticulously analyzed this information to identify true bug-fixing and bug-inducing commits. This approach ensures high data quality and reduces noise.

Details of these datasets are presented in Table \ref{tab:bug_statistics}. In this table, "SMALL" represents patches with fewer than five deleted lines, while "LARGE" represents patches with more than five deleted lines.
These datasets were further processed by Tang et al.~\cite{tang2023neural}, who identified the actual root-cause code lines representing the bugs, forming the final dataset used in our study.

\begin{table}[!t]
\centering
\caption{The statistics of the bugs and corresponding bug-fixing commits in three datasets}
\label{tab:bug_statistics}
\begin{adjustbox}{max width=\textwidth}
\begin{tabular}{lcccccc}
\toprule
\textbf{Dataset} & \textbf{Project} & \textbf{\#Bug-Fixing} & \textbf{\#bug-inducing} & \textbf{\#SMALL} & \textbf{\#LARGE} \\
\midrule
\multirow{6}{*}{Dataset1} 
& accumulo & 35 & 55 & 20 & 15 \\
& ambari & 38 & 44 & 17 & 21 \\
& hadoop & 53 & 57 & 28 & 25 \\
& lucene & 70 & 145 & 41 & 29 \\
& oozie & 45 & 50 & 23 & 22 \\
\cmidrule{2-6}
& \textbf{Total} & \textbf{241} & \textbf{351} & \textbf{129} & \textbf{112} \\
\midrule
\multirow{7}{*}{Dataset2} 
& jsoup & 63 & 63 & 35 & 28 \\
& fastjson & 222 & 222 & 144 & 78 \\
& verdictt & 53 & 53 & 11 & 42 \\
& closure-templates & 32 & 32 & 7 & 25 \\
& twilio-java & 39 & 39 & 14 & 25 \\
& ...(120 more projects) & 548 & 548 & 328 & 220 \\
\cmidrule{2-6}
& \textbf{Total} & \textbf{957} & \textbf{957} & \textbf{539} & \textbf{418} \\
\midrule
\multirow{6}{*}{Dataset3} 
& mockito & 32 & 53 & 13 & 19 \\
& joda-time & 23 & 27 & 12 & 11 \\
& commons-math & 85 & 111 & 44 & 41 \\
& total & 53 & 65 & 36 & 16 \\
& closure-compiler & 98 & 122 & 61 & 37 \\
\cmidrule{2-6}
& \textbf{Total} & \textbf{291} & \textbf{378} & \textbf{166} & \textbf{124} \\
\bottomrule
\end{tabular}
\end{adjustbox}
\end{table}

\subsection{State-of-The-Art Approaches}
For performance comparison, we adopted the state-of-the-art approaches used by Tang et al.~\cite{tang2023neural}, which include various machine learning algorithms: Random Forest (RF), Linear Regression (LR), Support Vector Machine (SVM), XGBoost (XGB), and K-Nearest Neighbors (KNN), as well as a deep learning-based approach Bi-LSTM. We also employed their state-of-the-art approach, which uses HAN as a heterogeneous graph neural network, referred to as Neural SZZ, as a baseline for comparison. For these approaches, we attempted to reproduce their approaches on the dataset. Our goal is to determine whether RC\_Detector exhibits superior performance compared to these state-of-the-art approaches. Since RC\_Detector mainly compares with the existing methods with the best performance, and the original paper does not disclose the specific hyperparameter settings of these methods, we directly used the performance results reported in the original paper for all methods except Neural SZZ. For Neural SZZ, we carefully reviewed the code disclosed by the paper's authors to confirm its completeness and consistency with the original paper description, and then performed comprehensive training on the original dataset and tried to reproduce it. Since the original author also did not give specific hyperparameters in the paper, we adjusted the parameters many times to approximate their results as closely as possible. Although the final metrics are slightly different from the original paper, the deviation is not large and may be caused by factors such as random seeds, implementation details, or version differences. For the sake of fairness and reproducibility, we still used the results reported in the original paper in subsequent comparisons.

\subsection{Evaluation Metrics}

To assess the effectiveness of the RC\_Detector model, we use two metrics proposed in previous studies: Recall@N and Mean First Rank (MFR) as indicators of the model's performance.

\subsubsection{Recall@N} 
Recall@N is a commonly used metric for evaluating recommendation systems and ranking models. This metric measures the model's ability to identify the true defects within the top N most likely defective code changes. The calculation formula is as follows:

\begin{equation}
Recall@N = \frac{{TP\;in\;TopN}}{{Total\;Actual\;Defects}} \times 100\% 
\end{equation}
where TP in Top N represents the number of true defects identified within the top N predicted results, and 
Total Actual Defects denotes the total number of actual defects. In this study, we set N=1,2,3. A higher Recall@N value indicates that the model is more effective in detecting defects within limited resources and time.

\subsubsection{Mean First Rank (MFR)} 
Mean First Rank (MFR) is another important metric for evaluating the performance of recommendation systems, used to measure the model's ability to rank the first correctly identified defect. Specifically, MFR represents the average rank position of all bug-inducing changes, with lower values indicating better performance. The formula is as follows:

\begin{equation}
MFR = \frac{1}{{\left| {Defects} \right|}}\sum\limits_{i = 1}^{\left| {Defects} \right|} {Ran{k_i}} 
\end{equation}
where Defects is the set of all identified defects, and ${Ran{k_i}}$ represents the rank of the $i$-th defect in the predicted results.

\subsection{Training Details} 
In this section, we detail the settings used for training the RC\_Detector model.

Previous researchers have primarily employed a ten-fold cross-validation protocol with critical time constraints~\cite{fushiki2011estimation}, ensuring that all changes in the test set chronologically follow those in the training set. This temporal validation strategy not only preserves the natural sequence of software development but also provides a rigorous and realistic assessment of the model's predictive performance in real-world scenarios, where future code changes are always predicted based on past data. Therefore, we also validated our model using ten-fold cross-validation. In this approach, all data is divided into ten parts, with each part used as a validation set while the remaining nine parts serve as the training set. The model is then trained and validated on these partitions, and the results are finally averaged.

For the embedding layer in the bug-fixing graph construction component, we use a pre-trained codeBERT model from the HuggingFace Transformer library to generate a 768-dimensional embedding for each code line. 
Based on the analysis of the impact of different learning rates on model performance during our experiments, we set the initial learning rate to 0.000005 and used the Adam optimizer for optimization. We set the number of attention heads to 8 based on experimental results. Finally, we incorporated layer normalization in the last layer to stabilize the training process and enhance the model's generalization ability.

The experimental environment is a computer equipped with an NVIDIA RTX A6000 GPU, 13th Gen Intel(R) Core(TM) i9-13900K, running on 22.04.1-Ubuntu OS. The programming language is Python 3.9 with torch-geometric 2.5.3, torch 2.4.0 and transformers 4.39.3 packages.
More detailed environmental information is also available in the GitHub readme~\cite{myreplication}, where we also give instructions on how to run each of our experiments, as well as a brief description of each code file.

\section{Experimental Results} \label{results}

\subsection{RQ1: Does our model really perform better than these state-of-the-art approches?}

\textbf{Motivation}: The purpose of this experiment is to validate the effectiveness of the RC\_Detector model in detecting the root causes within bug-fixing commits and to compare its performance against state-of-the-art approaches.

\textbf{Approach}: To assess the effectiveness of the RC\_Detector model, we compared its performance with RF, LR, SVM, XGB, KNN, Bi-LSTM, and Neural SZZ across Recall@1, Recall@2, Recall@3 scores, and MFR. For the state-of-the-art approaches, we reproduce their experiments and experimental settings. All experiments were conducted on the datasets mentioned in Section 4.1.

\begin{table}[!t]
\centering
\caption{Comparison of RC\_Detector and state-of-the-art approaches in terms of Recall@N, MFR}
\label{tab:RC_comparison}
\begin{tabular}{lcccc}
\toprule
\textbf{Approach} & \textbf{Recall@1} & \textbf{Recall@2} & \textbf{Recall@3} & \textbf{MFR} \\
\midrule
RF & 0.694 & 0.811 & 0.882 & 3.295 \\
LR & 0.701 & 0.813 & 0.872 & 3.541 \\
SVM & 0.714 & 0.806 & 0.869 & 3.215 \\
XGB & 0.718 & 0.811 & 0.867 & 3.133 \\
KNN & 0.677 & 0.792 & 0.860 & 2.773 \\
\midrule
Bi-LSTM & 0.656 & 0.746 & 0.820 & 3.448 \\
Neural SZZ & 0.779 & 0.841 & 0.886 & 2.425 \\
\textbf{RC\_Detector} & \textbf{0.813} & \textbf{0.900} & \textbf{0.929} & \textbf{1.799} \\
\bottomrule
\end{tabular}
\end{table}

\textbf{Results}: As shown in Table \ref{tab:RC_comparison}, the RC\_Detector model achieved superior performance in Recall@1, Recall@2, Recall@3 scores, and MFR across the whole dataset. On the combined dataset, RC\_Detector emerged as the best-performing model, with Recall@1, Recall@2, Recall@3, and MFR values of 0.813, 0.900, 0.929, and 1.799, respectively, surpassing the best state-of-the-art approach by 4.32\%, 7.06\%, 4.81\%, and 34.82\%. This indicates that RC\_Detector effectively captures code semantics for defect prediction. Recall@N reflects the percentage of true positives identified as the most likely bug-inducing commits, calculated as the ratio of the model's top-N true positives to the total number of true positives. A higher Recall@N value suggests better prioritization of critical defects.

RC\_Detector achieved a Recall@1 of 0.813, indicating that 81.3\% of the most critical bug-inducing commits were correctly identified, outperforming the best state-of-the-art approach by 4.32\%. When extending the measurement to the top 2 and top 3 predicted results, RC\_Detector reached 0.900 and 0.929, respectively, 
demonstrating consistent ability in capturing multiple high-risk commits, ensuring that even with multiple bug-inducing commits, they are accurately identified early. MFR measures how early the first correct prediction appears in the ranking list; the lower the MFR, the earlier the correct prediction appears. RC\_Detector's MFR was 1.799, 34.82\% lower than the best state-of-the-art approach, indicating that the model is not only accurate but also more efficient in bringing the most critical defect commits to the forefront. This efficiency is crucial for developers to address the most severe defects in practical applications promptly.

Upon reviewing the results, it is evident that traditional machine learning algorithms struggle to accurately predict defects using term frequency features, as this approach neglects the contextual relationships between code lines. Compared to ML state-of-the-art approaches, the deep learning approach Bi-LSTM performed relatively poorly, consistent with the findings of Wu et al.~\cite{9371393}. According to their research, simple text classification approaches perform better on clean datasets than specially designed deep learning approaches. The advantage of RC\_Detector over traditional approaches lies in its reduced reliance on manual feature engineering and its ability to better handle complex code semantic relationships, thereby improving prediction accuracy and reliability.

\textbf{Conclusion}: RC\_Detector generally outperforms state-of-the-art approaches, achieving superior results in Recall@N scores and MFR in all cases. RC\_Detector not only accounts for the differences in bug-fixing commits but also regulates information during the semantic propagation process, providing better performance than state-of-the-art approaches.

\subsection{RQ2: How does the RC\_Detector impact the performance in comparison to individual components?}

\begin{table}[!t]
\centering
\caption{The performance comparisons in ablation study}
\label{tab:ablation_study}
\begin{tabular}{lcccc}
\toprule
\textbf{Approach} & \textbf{Recall@1} & \textbf{Recall@2} & \textbf{Recall@3} & \textbf{MFR} \\
\midrule
Neural SZZ & 0.779 & 0.841 & 0.886 & 2.425 \\
Neural SZZ-GRU & 0.784 & 0.870 & 0.911 & 1.958 \\
RC\_Detector-g & 0.799 & 0.878 & \textbf{0.935} & \textbf{1.693} \\
RC\_Detector-h & 0.775 & 0.867 & 0.901 & 2.094 \\
\textbf{RC\_Detector} & \textbf{0.813} & \textbf{0.900} & 0.929 & 1.799 \\
\bottomrule
\end{tabular}
\end{table}

\textbf{Motivation}: This experiment focuses on two main components of the RC\_Detector model: the code semantic aggregation component and the cross-line semantic retention component. The objective of this experiment is to compare the performance of models built with individual components and demonstrate that the structural framework of RC\_Detector is superior to models composed of a single component. Meanwhile, to further examine the role of the cross-line semantic retention component, we conducted the experiment of Neural SZZ-GRU, which is the original Neural SZZ method plus the cross-line semantic retention component.

\textbf{Approach}:
To evaluate the effectiveness of each component, we compared RC\_Detector with its two variants: RC\_Detector-g and RC\_Detector-h. Each variant lacks a key design element. In RC\_Detector-g, we removed the code semantic aggregation component and fed the input directly into the cross-line semantic retention component. In RC\_Detector-h, we removed the cross-line semantic retention component and sent the output of the code semantic aggregation component directly to the RankNet model. Both RC\_Detector-g and RC\_Detector-h share the same graph construction process as RC\_Detector. We also conducted an experiment to compare the situations where the Neural SZZ method used GRU and did not use GRU. 

\textbf{Results}:
Table \ref{tab:ablation_study} compares the performance of RC\_Detector with its two variants, RC\_Detector-g and RC\_Detector-h, in identifying bug-inducing commits. The best results are highlighted in bold. Except for Recall@3 and MFR, RC\_Detector outperforms both variants across all metrics. In Recall@1, RC\_Detector surpasses RC\_Detector-g and RC\_Detector-h by 1.7\% and 4.9\%, respectively. Similarly, in Recall@2, it exceeds the variants by 2.5\% and 3.8\%. Although RC\_Detector does not achieve the best results in Recall@3 and MFR, the differences between its performance and the best results from RC\_Detector-g are minimal, with a decrease of only 0.6\% and 6\%, respectively, while it outperforms RC\_Detector-h by 3.1\% and 16.3\%. The performance differences can be attributed to the design of the model components. Unlike RC\_Detector, RC\_Detector-g only considers the semantic representation of individual code lines and applies gating to regulate this representation, neglecting the influence of context, which results in inferior performance on several metrics. For RC\_Detector-h, although it accounts for the influence of related code lines, its lack of regulating capability for propagated information prevents it from effectively retaining cross-line semantic relationships and critical information. That leads to its weaker performance compared to RC\_Detector. Additionally, we observed that the performance of RC\_Detector-g consistently surpasses that of RC\_Detector-h, which aligns with the findings of Tang et al.~\cite{tang2023neural}. The semantic representation of individual code lines plays a more crucial role in identifying bug-inducing commits than that of related code lines. The advantage of RC\_Detector over models composed of a single component lies in its ability to combine multiple components synergistically, thereby better capturing the contextual semantics of code lines and improving the accuracy of JIT prediction. However, compared to single-component models, RC\_Detector’s drawback is that it requires more computational resources and time. Meanwhile, the cross-line semantic retention component has also achieved significant improvements in the Neural SZZ method. Using GRU has increased by 0.6\%, 3.4\%, 2.8\%, and 23. 8\%, respectively, in various indicators compared to not using it.

\textbf{Conclusion}:
The experimental results show that the RC\_Detector model outperforms models composed of single components in terms of Recall@N and MFR. Thus, the structural framework of RC\_Detector offers greater potential for improvement than individual components. This makes RC\_Detector an effective approach for identifying the root causes of bugs, thereby enhancing JIT defect prediction. The results also show that the use of the GRU part is always much better than when it is not used, proving the importance of the cross-line semantic retention component.

\subsection{ RQ3:How do architectural choices impact RC\_Detector's performance? }

\subsubsection{Gating Mechanisms}
\
\par 
\noindent 
\textbf{Motivation}: The cross-line semantic retention component in RC\_Detector is designed to enhance the model's ability to integrate information. To evaluate the performance differences between RC\_Detector and other gating mechanisms in identifying root causes of defects, we tested various approaches.

\begin{table}[!t]
\centering
\caption{Comparison of RNN approaches on Recall@N, MFR}
\label{tab:rnn_comparison}
\begin{tabular}{lcccc}
\toprule
\textbf{Approach} & \textbf{Recall@1} & \textbf{Recall@2} & \textbf{Recall@3} & \textbf{MFR} \\
\midrule

RC\_Detector\_LSTM & 0.787 & 0.873 & 0.918 & \textbf{1.774} \\
RC\_Detector\_Hidden-Bias Simplified GRU & 0.804 & 0.882 & 0.918 & 1.909 \\
RC\_Detector\_Hidden-Only GRU & 0.784 & 0.873 & 0.906 & 2.051 \\
RC\_Detector\_Bias-Only GRU & 0.771 & 0.853 & 0.908 & 2.171 \\
RC\_Detector\_Transformer & 0.792 & 0.869 & 0.908 & 2.008 \\
\textbf{RC\_Detector\_GRU} & \textbf{0.813} & \textbf{0.900} & \textbf{0.929} & 1.799 \\
\bottomrule
\end{tabular}
\end{table}

\textbf{Approach}:
To determine whether GRU is suitable for the RC\_Detector model, we tested five different approaches: LSTM, three GRU variants, and Transformer. LSTM is an advanced Recurrent Neural Network (RNN)~\cite{6795963} that introduces forget, input, and output gates, effectively controlling the flow of information and selectively retaining or forgetting data. The GRU variants, proposed by Dey R. et al.~\cite{dey2017gate}, are modifications of GRU aimed at improving computational efficiency by simplifying parameters while maintaining a high level of processing capability. These three GRU variants are named Hidden-Bias Simplified GRU, Hidden-Only GRU, and Bias-Only GRU, according to the specific elements they utilize for gate computations. Hidden-Bias Simplified GRU uses both the previous hidden state and bias, Hidden-Only GRU relies solely on the previous hidden state, and Bias-Only GRU uses only the bias term. Transformer is renowned for its attention mechanism, which allows the model to simultaneously weigh the importance of different parts of the input sequence.
We applied these five approaches within the cross-line semantic retention component to compare the impact of different gating mechanisms on the RC\_Detector model.

\textbf{Results}:
Table \ref{tab:rnn_comparison}  compares the results using GRU with four other RNN approaches (LSTM, Hidden-Bias Simplified GRU, Hidden-Only GRU, and Bias-Only GRU) and Transformer.
 In terms of Recall@1, Recall@2, Recall@3, and MFR, GRU outperforms LSTM, Hidden-Bias Simplified GRU, Hidden-Only GRU, and Bias-Only GRU across most results in the dataset. Specifically, in terms of average Recall@1, GRU's performance exceeds that of LSTM, Hidden-Bias Simplified GRU, Hidden-Only GRU, and Bias-Only GRU by 3.30\%, 1.12\%, 3.70\%, and 5.45\%, respectively. For Recall@2, GRU outperforms LSTM, Hidden-Bias Simplified GRU, Hidden-Only GRU, and Bias-Only GRU by 3.09\%, 2.04\%, 3.09\%, and 5.51\%, respectively. In terms of Recall@3, GRU surpasses LSTM, Hidden-Bias Simplified GRU, Hidden-Only GRU, and Bias-Only GRU by 1.20\%, 1.20\%, 2.54\%, and 2.31\%, respectively. Regarding MFR, GRU achieves a lower MFR compared to Hidden-Bias Simplified GRU, Hidden-Only GRU, and Bias-Only GRU by 6.11\%, 14.01\%, and 20.68\%, although it is 1.39\% higher than LSTM.

Surprisingly, the Transformer did not achieve the best results; compared to the optimal performance of GRU, it performed lower by 2.57\%, 3.49\%, 2.25\%, and 10.42\% across various metrics. Its performance was also only mediocre when compared to the other four RNN approaches.

Compared to GRU, LSTM introduces forget, input, and output gates, forming a more complex gating mechanism. However, the additional gating parameters might increase the risk of overfitting, leading to suboptimal results. On the other hand, the parameter simplifications in the GRU variants may weaken the model's expressiveness, reducing its ability to capture crucial semantic information in more complex and diverse code change scenarios, thereby leading to a decline in performance. As for the Transformer, it may also have suffered from overfitting issues, and its higher sensitivity to hyperparameter tuning likely required more precise adjustments to achieve optimal results. Thus, GRU proves to be an effective architecture for handling complex code change tasks, particularly demonstrating significant advantages in Recall metrics. Overall, GRU outperforms the other approaches evaluated in this study (LSTM, Hidden-Bias Simplified GRU, Hidden-Only GRU, Bias-Only GRU, and Transformer) across the Recall@1, Recall@2, Recall@3, and MFR metrics.

\textbf{Conclusion}:
The results indicate that GRU outperforms LSTM, Hidden-Bias Simplified GRU, Hidden-Only GRU, Bias-Only GRU, and Transformer in terms of Recall@1, Recall@2, Recall@3, and MFR. This suggests that GRU can effectively be used in the cross-line semantic retention component to filter critical contextual information in code changes, offering superior performance compared to other gating approaches.
\subsubsection{Attention Mechanisms}
\
\par 
\noindent 
\textbf{Motivation}: Attention mechanisms play a crucial role in enabling models to focus on the most relevant parts of the input data when making predictions. In the context of the RC\_Detector model, integrating an effective attention mechanism can significantly enhance the model's ability to capture complex relationships and dependencies within heterogeneous graph data. To evaluate and improve the information aggregation capability of RC\_Detector, we explore and compare different attention mechanisms to understand their impact on the model’s performance in identifying root-cause errors.

\textbf{Approach}:
To assess the applicability of various attention mechanisms within the RC\_Detector framework, we experimented with five mechanisms, each employing a different computation approach:

\textbf{Additive Attention}~\cite{bahdanau2014neural}: This mechanism adds query and key vectors through a feedforward network and applies a nonlinear function to compute attention weights, capturing correlations between input vectors.

\textbf{Graph Attention Network (GAT)}~\cite{velivckovic2017graph}:GAT is a graph neural network that computes node representations by adaptively aggregating features from neighboring nodes through an attention mechanism. GAT only needs information about the first-order neighbor nodes, which allows it to handle a wider range of graph data. GAT calculates attention weights by performing a linear combination of each pair of node features and applying a LeakyReLU activation function to generate initial weights.

\textbf{Cosine Similarity Attention}: This approach determines attention weights by calculating the cosine similarity between node feature vectors.

\textbf{Scaled Dot-Product Attention}~\cite{vaswani2017attention}: Scaled dot-product attention computes attention weights by calculating the dot product of query and key vectors, followed by scaling and normalization.

\textbf{Gaussian Kernel Function}~\cite{chen2021skyformer}: This mechanism computes attention weights by calculating the Euclidean distance between the query and key vectors, squaring the result, and applying a Gaussian function.

We integrated these attention mechanisms into the RC\_Detector model and evaluated their performance using several metrics, including Recall@1, Recall@2, Recall@3, and MFR.

\begin{table}[!t]
\centering
\caption{Comparison of Scaled Dot-Product Attention and Other Attention Mechanisms}
\label{tab:attention_comparison}
\begin{adjustbox}{max width=\textwidth}
\begin{tabular}{lcccc}
\toprule
\textbf{Approach} & \textbf{Recall@1} & \textbf{Recall@2} & \textbf{Recall@3} & \textbf{MFR} \\
\midrule
RC\_Detector\_Additive Attention & 0.790 & 0.874 & 0.917 & 2.120 \\
RC\_Detector\_Graph Attention Network & 0.802 & 0.874 & 0.908 & 2.009 \\
RC\_Detector\_Cosine Similarity Attention & 0.784 & 0.866 & 0.905 & 2.017 \\
RC\_Detector\_Gaussian Kernel Function & 0.797 & 0.869 & 0.909 & 2.620 \\
\textbf{RC\_Detector\_Scaled Dot-Product Attention} & \textbf{0.813} & \textbf{0.900} & \textbf{0.929} & \textbf{1.799} \\
\bottomrule
\end{tabular}
\end{adjustbox}
\end{table}

\textbf{Results}:
Table \ref{tab:attention_comparison} presents our experimental results. By comparing the performance of these attention mechanisms, it is evident that Scaled Dot-Product Attention consistently excels across all metrics. It achieved Recall@1, Recall@2, and Recall@3 values of 0.813, 0.900, and 0.929, respectively, with an MFR of only 1.799. These values consistently ranked the highest compared to other techniques, outperforming the lowest values by 3.62\%, 3.95\%, 2.62\%, and 45.64\%, respectively.

In contrast, the Graph Attention Network approach, while slightly less effective, still performed well with Recall@1, Recall@2, Recall@3, and MFR values of 0.802, 0.874, 0.908, and 2.009, respectively. These results were 1.30\%, 2.97\%, 2.25\%, and 10.48\% lower than those of Scaled Dot-Product Attention but were still superior to the other three approaches in terms of Recall@1 and MFR.
Additive Attention achieved Recall@1, Recall@2, Recall@3, and MFR values of 0.790, 0.874, 0.917, and 2.120, respectively. It performed relatively well in Recall@3, reaching 0.917, which was only 1.25\% lower than the best result, and outperformed Graph Attention Network, Cosine Similarity Attention, and Gaussian Kernel Function by 1.02\%, 1.34\%, and 0.83\%, respectively. However, its MFR of 2.120 was slightly higher than that of the other approaches. When using Cosine Similarity Attention, the results were the lowest in terms of Recall, with values of 0.784, 0.866, and 0.905, but the MFR remained at a mid-level of 2.017. Notably, although the Gaussian Kernel Function showed balanced performance across all Recall metrics, with values of 0.797, 0.869, and 0.909, its MFR of 2.620 was the worst among all approaches, 19.07\% lower than the next worst result.

These findings collectively demonstrate that Scaled Dot-Product Attention achieves superior results on the experimental dataset compared to other attention mechanisms. Additionally, RC\_Detector consistently exhibits significant improvements over state-of-the-art approaches across most evaluation metrics, confirming its adaptability to different attention-weight computation approaches.

\textbf{Conclusion}:
In summary, our results confirm the effectiveness of the RC\_Detector model architecture in accurately capturing the hidden semantics of code lines, thereby improving the accuracy of identifying bug-inducing changes. Notably, when Scaled Dot-Product Attention is used as the attention mechanism, the RC\_Detector model achieves the most commendable performance. 
It is worth highlighting that the model’s structure remains robust and effective, regardless of the specific attention computation approach integrated into the components.

\begin{table}[!t]
\centering
\caption{Comparison of Recall@N, MFR for Different Learning Rates and Number of Attention Heads}
\label{tab:learning_rate_comparison}
\begin{adjustbox}{max width=\textwidth}
\begin{tabular}{cccccc}
\toprule
\textbf{Learning Rate} & \textbf{Head Num} & \textbf{Recall@1} & \textbf{Recall@2} & \textbf{Recall@3} & \textbf{MFR} \\
\midrule
\multirow{2}{*}{5.00E-06} & 8  & \textbf{0.813} & \textbf{0.900} & \textbf{0.929} & \textbf{1.799} \\
                          & 16 & 0.804 & 0.878 & 0.908 & 2.611 \\
\midrule
\multirow{2}{*}{1.00E-06} & 8  & 0.793 & 0.875 & 0.920 & 1.949 \\
                          & 16 & 0.799 & 0.891 & 0.923 & 1.980 \\
\midrule
\multirow{2}{*}{1.00E-05} & 8  & 0.722 & 0.814 & 0.866 & 2.394 \\
                          & 16 & 0.782 & 0.861 & 0.908 & 2.804 \\
\bottomrule
\end{tabular}
\end{adjustbox}
\end{table}

\subsection{RQ4: How does our model perform under different hyperparameters?}

\textbf{Motivation}: As discussed earlier in Section 4.4, the choice of hyperparameters can significantly impact the model's performance. In this context, we focus on two key hyperparameters: the learning rate and the number of attention heads~\cite{michel2019sixteen}. The learning rate determines the speed at which the model adjusts its parameters during training, with an inappropriate learning rate potentially leading to slow convergence or poor generalization. On the other hand, the number of attention heads affects the model's ability to capture different aspects of the data in parallel, influencing overall model complexity and performance.

\textbf{Approach}:
To explore the impact of these hyperparameters, we conducted experiments using three different learning rates (5.00E-06, 1.00E-06, and 1.00E-05) and evaluated the model’s performance with both 8 and 16 attention heads for each learning rate setting. We measured the model’s performance using metrics such as Recall@1, Recall@2, Recall@3, and Mean First Rank (MFR) and employed ten-fold cross-validation for the experiments.

\textbf{Results}:
The results are summarized in the provided table \ref{tab:learning_rate_comparison}. The experiments revealed that the model performed best when the learning rate was 5.00E-06 and 8 attention heads were used, achieving a Recall@1 of 0.813, Recall@2 of 0.900, Recall@3 of 0.929, and an MFR of 1.799. As the learning rate was decreased to 1.00E-06, the model’s performance declined, with Recall@1 dropping to 0.793, Recall@2 to 0.875, Recall@3 to 0.920, and MFR increasing to 1.949, representing decreases of 2.44\%, 2.89\%, 0.97\%, and 8.35\%, respectively, compared to the optimal results. However, when the number of attention heads was increased to 16, the lower learning rate resulted in some improvement, with Recall@2, Recall@3, and MFR increasing by 1.51\%, 1.63\%, and 31.87\%, respectively, despite a 0.68\% drop in Recall@1.

When the learning rate was increased to 1.00E-05, the performance across all metrics deteriorated significantly, with the lowest performance occurring with 8 attention heads, where Recall@1, Recall@2, Recall@3, and MFR were 0.7215, 0.8140, 0.8660, and 2.3943, respectively, marking declines of 12.63\%, 10.61\%, 7.23\%, and 33.11\% from the best results. This clearly indicates that the higher learning rate hindered effective convergence during training, negatively affecting the model’s generalization ability during testing.

While theoretically, increasing the number of attention heads should enable the model to focus on more feature dimensions, the experimental results showed a performance decline when the number of attention heads was increased from 8 to 16 at a learning rate of 5.00E-06, with decreases of 1.01\%, 2.47\%, 2.23\%, and 31.10\% for Recall@1, Recall@2, Recall@3, and MFR, respectively. However, at a learning rate of 1.00E-06, aside from a 1.57\% decline in MFR, the other metrics improved by 0.71\%, 1.87\%, and 0.32\%. A similar trend was observed at a learning rate of 1.00E-05, where the metrics improved by 8.40\%, 5.76\%, and 4.89\%, with MFR decreasing by 14.62\%.

\textbf{Conclusion}:
Based on these experimental results, we conclude that the optimal hyperparameters for this experimental setup are a learning rate of 5.00E-06 and 8 attention heads. This configuration provides the best balance between model complexity and performance, allowing the model to effectively capture and generalize information from the data while maintaining high recall rates and a low MFR.

\begin{table}[!t]
\centering
\caption{Comparison of different embedding layers in RC\_Detector}
\label{tab:embedding_layer_comparison}
\begin{adjustbox}{max width=\textwidth}
\begin{tabular}{lcccc}
\toprule
\textbf{Embedding Layer} & \textbf{Recall@1} & \textbf{Recall@2} & \textbf{Recall@3} & \textbf{MFR} \\
\midrule

Graphcodebert & 0.786 & 0.871 & 0.917 & 1.965 \\
CodeT5 & 0.650 & 0.762 & 0.832 & 2.828 \\
UniXcoder & 0.772 & 0.863 & 0.902 & 2.116 \\
\textbf{CodeBERT} & \textbf{0.813} & \textbf{0.900} & \textbf{0.929} & \textbf{1.799} \\
\bottomrule
\end{tabular}
\end{adjustbox}
\end{table}

\subsection{RQ5: How does the performance of the RC\_Detector model vary under different embedding layers?}
\textbf{Motivation}: The choice of embedding layer has a crucial impact on the performance of neural network models, especially in tasks involving code understanding and generation. To optimize the performance of the RC\_Detector model, we evaluated four different embedding layers—CodeBERT~\cite{feng2020codebert}, Graphcodebert~\cite{guo2020graphcodebert}, CodeT5~\cite{wang2021codet5}, and UniXcoder~\cite{guo2022unixcoder}—to explore their effects on the model’s performance in code processing tasks.

\textbf{Approach}:
We conducted experiments using these four different embedding layers within the RC\_Detector model and compared the results. CodeBERT is a bimodal pre-trained model for programming languages and natural language, utilizing masked language modeling (MLM) and replaced token detection (RTD) tasks to learn a joint representation of code and natural language~\cite{feng2020codebert}. Graphcodebert, on the other hand, is pre-trained based on the data flow in code structures, capturing semantic dependencies within the code and making it suitable for tasks that require understanding code structure~\cite{guo2020graphcodebert}. CodeT5~\cite{wang2021codet5} employs an encoder-decoder architecture, supporting both code understanding and generation tasks, and introduces identifier-aware pre-training tasks to enhance the understanding of code semantics. UniXcoder~\cite{guo2022unixcoder}is a unified cross-modal pre-trained programming language model that supports code understanding and generation tasks, particularly suitable for multi-task learning and aligning natural language with programming languages.

\textbf{Results}:
The experimental results are shown in Table \ref{tab:embedding_layer_comparison}, providing metrics for the four different embedding layers. The RC\_Detector model with the CodeBERT embedding layer performed the best, achieving Recall@1, Recall@2, and Recall@3 values of 0.813, 0.900, and 0.929, respectively, with an MFR of 1.799. In comparison, Graphcodebert’s performance was slightly lower, with corresponding metrics of 0.786, 0.871, 0.917, and 1.965. CodeT5 performed the worst, with all Recall metrics lower than those of the other embedding layers, achieving 0.650, 0.762, 0.832, and an MFR as high as 2.828, which is 36.39\% lower than the best result. UniXcoder’s performance fell between that of Graphcodebert and CodeT5, with Recall@1, Recall@2, Recall@3, and MFR values of 0.772, 0.863, 0.902, and 2.116, respectively.

These results clearly indicate that the initial semantic representation provided by the embedding layer plays a critical role in determining the effectiveness of the RC\_Detector model. Among the tested options, CodeBERT emerged as the best choice, delivering optimal performance across all metrics. This also suggests that, for similar tasks, careful consideration of the embedding layer can significantly improve the model’s task performance.

\textbf{Conclusion}:
Overall, CodeBERT is the most suitable embedding layer for use within the RC\_Detector model, offering the best performance across various metrics. Graphcodebert and UniXcoder also show some competitive strengths, particularly in adapting to specific tasks. However, CodeT5’s poor performance in this experiment suggests that it may not be well-suited to the specific requirements of the RC\_Detector model. Based on these results, we prioritize the CodeBERT embedding layer to achieve optimal model performance.

\begin{table}[!t]
\centering
\caption{The performance comparisons in identifying bug-inducing commits}
\label{tab:identifying bug-inducing_comparison}
\begin{adjustbox}{max width=\textwidth}
\begin{tabular}{lcccc}
\toprule
\textbf{Approach} & \textbf{Precision} & \textbf{Recall} & \textbf{F1-score} \\
\midrule
        BSZZ & 0.376 & 0.730 & 0.496 \\ 
        AG-SZZ & 0.348 & 0.604 & 0.441 \\ 
        MA-SZZ & 0.319 & 0.543 & 0.401 \\ 
        RA-SZZ & 0.333 & 0.466 & 0.388 \\ 
\midrule
        NeuralSZZ@1 & \textbf{0.834} & 0.598 & 0.698 \\ 
        NeuralSZZ@2 & 0.728 & 0.635 & 0.678  \\ 
        NeuralSZZ@3 & 0.685 & 0.667 & 0.676 \\ 
\midrule
       RC\_Detector@1 & 0.817 & 0.677 &\textbf{0.740} \\ 
    RC\_Detector@2 & 0.710 & 0.756 & 0.732 \\ 
        RC\_Detector@3 & 0.649 & \textbf{0.798} & 0.715 \\ 
\bottomrule
\end{tabular}
\end{adjustbox}
\end{table}

\subsection{RQ6: How does the performance of the RC\_Detector model in identifying bug-inducing commits?}

\textbf{Motivation}:To explore whether deleting lines ranked by RC\_Detector can enhance the ability of SZZ to identify bug-inducing commits, we conducted this experiment using the same settings as NeuralSZZ.

\textbf{Approach}:
We examined the top 1, 2, and 3 lines ranked by RC\_Detector as most likely to be the root cause of the bug to see if they were indeed related to the bug-inducing commits. We calculated the precision, recall, and F1-score to comprehensively evaluate the model's performance.

\textbf{Results}:
Experimental results show that RC\_Detector performs significantly better than the previous static SZZ algorithm at detecting bug-inducing commits in table \ref{tab:identifying bug-inducing_comparison}. RC\_Detector also outperforms all NeuralSZZ methods in all metrics except Precision. Despite a decrease in Precision, our method achieves a significant improvement in Recall. In particular, RC\_Detector improves the recall of the first, second, and third lines by 13.21\%, 19.06\%, and 19.64\%, respectively. Since recall and precision are equally important, we use the F1-score as the main evaluation metric to avoid bias. The F1-score can measure whether the improvement in recall rate outweighs the decrease in precision. Compared to NEURALSZZ, RC\_Detector achieves a better balance between precision and recall. This is particularly evident in the F1-score, where RC\_Detector's F1-scores for the top 1, top 2, and top 3 lines are 0.740, 0.732, and 0.715, respectively, which are 6.01\%, 7.96\%, and 5.77\% higher than the baseline method. Therefore, we believe that RC\_Detector is more effective than the previous SZZ algorithm in detecting bug-inducing commits.

\textbf{Conclusion}:
RC\_Detector captures the semantic features of code lines through dynamic semantic modeling using a heterogeneous graph neural network, which has convincing advantages over the previous static method that only considers syntactic features. At the same time, it uses the double gating mechanism of GRU to retain the semantic information learned earlier, thereby solving the problem of homogenization of code line representations in deep networks and achieving better results in F1-score than NeuralSZZ. Overall, the RC\_Detector model outperforms the traditional NeuralSZZ method in identifying bug-inducing commits. In particular, RC\_Detector demonstrates superior performance in balancing precision and recall. Therefore, RC\_Detector has potential application value in the actual software development and maintenance process, helping development teams more effectively locate and fix defects and improve software quality.

\begin{table}[!t]
\centering
\caption{The performance comparisons in cross-project prediction}
\label{tab:performance_comparison}
\begin{adjustbox}{max width=\textwidth}
\begin{tabular}{lcccc}
\toprule
\textbf{Approach} & \textbf{Recall@1} & \textbf{Recall@2} & \textbf{Recall@3} & \textbf{MFR} \\
\midrule
RF         & 0.697 & 0.783 & 0.834 & 2.866 \\ 
LR         & 0.643 & 0.834 & 0.859 & 2.528 \\ 
SVM        & 0.681 & 0.789 & 0.866 & 2.630 \\ 
XGB        & 0.675 & 0.802 & 0.853 & 2.318 \\ 
KNN        & 0.675 & 0.770 & 0.847 & 3.369 \\
\midrule
Bi-LSTM    & 0.541 & 0.746 & 0.820 & 3.448 \\ 
NeuralSZZ  & 0.786 & 0.860 & 0.891 & 2.197 \\ 
\midrule
\textbf{RC\_Detector} & \textbf{0.796} & \textbf{0.866} & \textbf{0.898} & \textbf{2.0} \\ 
\bottomrule
\end{tabular}
\end{adjustbox}
\end{table}

\subsection{RQ7: How does the performance of the RC\_Detector model in cross-project settings?}

\textbf{Motivation}: Cross-project scenarios involve training a model using data from one or more projects and testing it in a different project. This setting is commonly used to assess the generalisability and reliability of a model across different code bases. Evaluating the performance of RC\_Detector in this context is critical for determining its suitability for projects that lack sufficient historical data.

\textbf{Approach}:In order to evaluate the cross-project performance of RC\_Detector and compare it with NeuralSZZ, we used the same settings as NeuralSZZ for the experiment:

\textbf{Train data:} DATASET2 and DATASET3 were used for training.

 \textbf{Test data:}  DATASET1 was selected as the test set due to its high-quality annotations and well-documented bug fixes committed.

\textbf{Results}: The experimental results are shown in Table \ref{tab:performance_comparison}. Results show that RC\_Detector achieved Recall@1: 0.796, Recall@2: 0.866, Recall@3: 0.898, and MFR: 2.0 in all metrics. Although these improvements are gradual compared to NeuralSZZ, they still show that RC\_Detector performs well in cross-project experiments. 

\textbf{Conclusion}:
The RC\_Detector model shows enhanced performance in the cross-project setting compared to the NeuralSZZ baseline. It is able to make predictions across different projects, which emphasizes its generalisability and reliability. RC\_Detector also has potential applicability even in the case of limited project-specific historical data. Since the number of data in the training set in the cross-project experiment is significantly reduced compared to the cross-validation, it is a disadvantage for RC\_Detector, which has more parameters, and its improvement in each metric is not as good as cross-validation.

\section{Threats to Validity} \label{Threats}
\subsection{Internal Validity} 
The main threat to internal validity is the correctness of the NSZZ implementation and the reproduction of state-of-the-art methods. To mitigate this, we conducted a thorough review of the NSZZ source code, comparing it to the detailed workflow and pseudocode provided in the original paper. Furthermore, when reproducing other baseline models, we used their publicly available code. However, there remains a threat to the accuracy of the labels in our test datasets. Despite the quality checks we have performed on the datasets, it is still possible that they are not entirely accurate. Because the labels in the test data sets contain noise, there is a potential threat to the internal validity of the experimental results.
\subsection{External Validity} 
External validity refers to the generalisability of the proposed RC detector model. In this study, we used three widely used bug-fixing datasets to evaluate our approach. Although these datasets are considered to be of high quality, a potential limitation is that they contain a relatively small number of bugs, which may not fully capture the variability present in more diverse or larger software projects. Therefore, our results may not be fully representative of all scenarios or project types that could benefit from RC\_Detector. However, the diversity of the selected datasets and the consistency of our observations strengthen our confidence in the broader applicability of the proposed model.
\subsection{Construct Validity} 
Construct validity refers to the evaluation metrics used in the RC\_Detector model for JIT defect prediction techniques. We use Recall@N to evaluate the performance of the model and Mean First Rank to evaluate the cost-effectiveness of the JIT model. to measure the effectiveness and cost-effectiveness of JIT predictions. Recall@N quantifies the proportion of lines of code identified as actually causing bugs that are among the top N lines of code, while Mean First Rank reflects the efficiency with which the model detects these bug-causing commits. These metrics are commonly used in software engineering research, which reduces the threat to the construct validity of our work.

\section{Conclusions and Future Work}
In this paper, we proposed the RC\_Detector model to enhance the accuracy of Just-In-Time (JIT) defect prediction. RC\_Detector consists of three main components: the bug-fixing graph construction component, the code semantic aggregation component, and the cross-line semantic retention component. First, we preprocess the bug-fixing commit datasets into heterogeneous graphs using the bug-fixing graph construction component. These graphs are then sequentially input into the code semantic aggregation component. The code semantic aggregation component learns the hidden semantic representation of target code lines by aggregating the semantics of code lines directly related to the target based on the constructed heterogeneous graph. This newly acquired semantic representation, along with the old semantic representation, is then passed to the cross-line semantic retention component. In this component, the attenuation gate and reinforcement gate are derived from the old and new semantic representations to regulate the information accordingly. Through the synergistic operation of these three main components, the RC\_Detector model more accurately captures and propagates semantic information between code lines, thereby improving the effectiveness of JIT defect prediction.

Our experiments, conducted on a dataset comprising 675 bug-fixing commits from 87 open-source projects, demonstrate that the RC\_Detector model outperforms state-of-the-art approaches. Specifically, it achieves improvements of 4.32\%, 7.06\%, 4.81\%, and 34.82\% over the best state-of-the-art approaches in Recall@1, Recall@2, Recall@3, and MFR, respectively, validating its effectiveness in enhancing the precision of JIT defect prediction.

In the future, we plan to denoise the dataset to enable more robust training of the model and to evaluate our model on a larger number of high-quality bug-fixing datasets to expand our research.


\bibliographystyle{unsrt}
\bibliography{ref}

\end{document}